# Sub-nanosecond, time-resolved, broadband infrared spectroscopy using synchrotron radiation

R.P.S.M. Lobo,[1,2] J.D. LaVeigne,[2] D.H. Reitze,[2] D.B. Tanner[2] and G. L. Carr[1]

[1] National Synchrotron Light Source, Brookhaven National Laboratory, Upton NY 11973-5000
[2] Department of Physics, University of Florida, Gainesville FL 32611

A facility for sub-nanosecond time-resolved (pump-probe) infrared spectroscopy has been developed at the National Synchrotron Light Source of Brookhaven National Laboratory. A mode-locked Ti:sapphire laser produces 2 ps duration, tunable near-IR pump pulses synchronized to probe pulses from a synchrotron storage ring. The facility is unique on account of the broadband infrared from the synchrotron, which allows the entire spectral range from 2 cm$^{-1}$ (0.25 meV) to 20,000 cm$^{-1}$ (2.5 eV) to be probed. A temporal resolution of 200 ps, limited by the infrared synchrotron-pulse duration, is achieved. A maximum time delay of 170 ns is available without gating the infrared detector. To illustrate the performance of the facility, a measurement of electron-hole recombination dynamics for an HgCdTe semiconductor film in the far- and mid infrared range is presented.

## I. INTRODUCTION

Synchrotron radiation is the light emitted by highly relativistic electrons (or other charged particles) as they transit the magnetic fields used to guide them along a closed orbit. The emitted radiation extends from microwaves to the hard x-ray region, and therefore includes the entire spectral range conventionally used for studies of the optical properties of solids. Numerous beamlines that are dedicated to infrared spectroscopy have been developed at synchrotron radiation facilities throughout the world.[1-4] Most were designed to exploit the very high brightness of infrared synchrotron radiation for microscopy and other throughput-limited techniques. A few have been instrumented for the far-infrared, where the synchrotron source offers both a brightness and power advantage over alternative spectroscopic sources.[5] Until recently, most studies with infrared synchrotron radiation utilized the high brightness of the source. Grazing incidence methods (such as for surface science studies),[6] infrared microscopy,[7,8] and ellipsometry[9] are examples of techniques that substantially benefit from the brightness of infrared synchrotron radiation.

A key difference between the synchrotron infrared source and a conventional thermal source is that the light is pulsed, with typically sub-nanosecond duration. These pulses are a direct consequence of the manner in which a synchrotron operates. The accelerating fields inside a radio frequency (RF) cavity restore the energy lost by the orbiting electrons through radiation emission. Because the RF field oscillates, only electrons arriving at a particular time receive the proper acceleration, leading to electron bunching. A typical bunch length is a few cm, leading to pulse durations of about 100 ps.

This temporal structure of the light pulses is widely underutilized, despite the fact that there are many phenomena to be investigated with time-resolved infrared spectroscopy in the nanosecond range. Recombination dynamics of the photo-generated electron-hole plasma in semiconductors can have characteristic time scales of tens of nanoseconds[10,11] as do some intersubband excitations in quantum wells.[12] In conventional superconductors, Cooper pairs can be broken by light absorption,[13] with the subsequent recombination into pairs occurring relatively slowly (∼ ns) for some conventional superconductors.[14] In the high-$T_c$ superconductors, pair-recombination dynamics may reveal information about the nature of the superconducting state. In addition, contradictory results have been obtained by infrared spectroscopy for the quasiparticle recombination time, with values ranging from picoseconds[15] to milliseconds[16] reported. Broadband, time-dependent measurements might resolve these differences. In the underdoped cuprates, pair breaking studies may provide fundamental information on the nature of the pseudogap state. The insulating phases of high-$T_c$ superconductors show a photoinduced conductivity effects.[17] The dynamics of this excitation may help understand the mechanism of the normal state conductivity. As an alternative to chemical doping, conducting polymers can be photodoped, creating polaron and soliton states.[18] In most cases, the decay from the excited state occurs on picosecond scales.[19] However in some cases, a slower response in the nanosecond range is present.[20-22]

Measurements using infrared synchrotron radiation to investigate the time dependence of photoexcitations in GaAs,[23] Hg$_x$Cd$_{1-x}$Te,[24,25] and superconducting Pb[26] have already proven the value of broadband infrared pump-probe measurements in the sub-nanosecond range. Here we provide a detailed discussion of the newly developed sub-nanosecond time-resolved spectroscopy facility at the National Synchrotron Light Source (NSLS) of Brookhaven National Laboratory. The facility makes use of the broadband infrared pulses from the VUV ring, spans the spectral range from 2 cm⁻¹ (250 μeV, 5 mm) to 20,000 cm⁻¹ (2.5 eV, 500 Å), and achieves a time resolution approaching 100 ps.

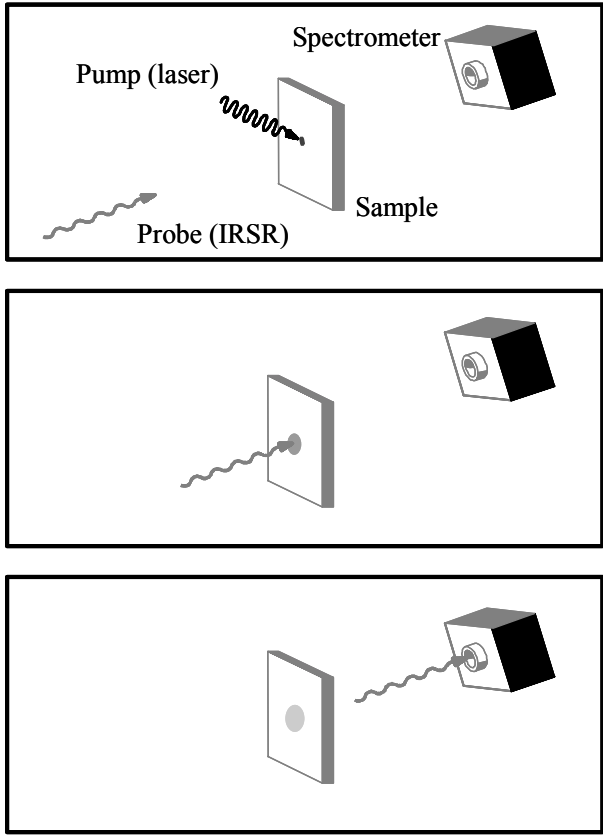

FIG. 1. Schematics of pump-probe experiments. The pump pulse (Ti:sapphire laser at U12IR) excites the sample, the probe pulses (synchrotron at U12IR) arriving a certain time (Δt) later analyses the sample sending to the detector a snapshot of the system as it was Δt after the excitation.

In Section 2 of this paper, we describe the pump-probe technique and survey pulsed sources of infrared. We also discuss the special considerations that are required for the implementation of pump-probe, far infrared spectroscopy. In section 3 we describe details of the laser synchronization, the optical delivery system and the spectrometers. Section 4 addresses the collection of photoinduced spectra. In section 5, we present a study of the dynamics of electron-hole recombination in $Hg_{1-x}Cd_xTe$ (MCT) films in the far- and mid-infrared regions to illustrate the capabilities of our system.

## II. PUMP-PROBE

### II.1 The technique

Time-resolved spectroscopy determines a material's spectral properties at various instants of time after some stimulus has been applied. There are two common approaches for achieving time resolution: (i) signal gating by sampling the output of a fast responding photodetector or (ii) probing with a short duration pulse of light. Although detectors with a fast response exist for many portions of the spectrum, few are available that operate at long wavelengths.[27,28] For example, the most sensitive bolometric detectors for the far infrared have response times of about 1 ms. Consequently, achieving high temporal resolution by gating the detector is simply unworkable. Even when fast detectors are available, the detector noise increases as $(B)^{1/2}$ where $B$ is the detector's response bandwidth. Thus the signal-to-noise for a detector with $B = 1$ GHz (for achieving 1 ns time resolution) will be 100 times poorer than for a similar detector having $B = 100$ kHz (as is typical for infrared spectrometers).

Pump-probe spectroscopy — an established technique for obtaining high temporal resolution of repeatable, photo-stimulated events[29] — can overcome these limitations. As illustrated in Fig. 1, a pump pulse excites the sample and a probe pulse, arriving at a time Δt after the pump, analyzes the sample's response at a time Δt into its relaxation process. The combination of pump-and-probe can be repeated at a high rate, and as long as a constant delay time Δt is maintained, the probe queries the system in the same state every time. A complete spectrum can be measured for this particular delay, representing a "snapshot" of the sample's state. This process is repeated for a wide range of Δt values to build up a complete time history of the sample's relaxation dynamics.

The duration of the pump and probe pulses, along with the impulse response function of the sample, determine the measured temporal response $S(\Delta t)$ in a pump-probe experiment. When the sample's response is linear:

$$S(\Delta t) = \int_{-\infty}^{+\infty} dt' \int_{-\infty}^{t'} dt'' \, I_{probe}(t' + \Delta t) \, I_{pump}(t'') \, G(t'') \quad (1)$$

where $I_{pump}(t)$ and $I_{probe}(t)$ are respectively the temporal intensity profiles of the pump and probe pulses and $G(t)$ is the sample's impulse response function (the quantity of interest). Setting $G(t) = \delta(t)$ yields the the cross correlation of the pump and probe pulses, which defines the minimum temporal resolution. Note that achieving a high temporal resolution is independent of the sensitivity and response time of spectrometer and detector. Also, Eq. 1 assumes sample excitation only by the pump pulse. Typically, this is achieved by selecting the probe pulse to be much less intense than the pump pulse, but may also be achieved by limiting the probe to a spectral range below some photon energy threshold.

### II.2 Pulsed IR Sources

Pump-probe measurements can be performed using a variety of excitation methods, including electrical current, electric field, magnetic field, or (most commonly) a light pulse. An ideal facility for studying all potentially interesting scientific problems would have a completely tunable pump source that provides adequate energy density (fluence) to create the desired density of excitations in the sample. Pulse duration and repetition frequencies (PRFs) would also be variable to match a specimen's relaxation time. The probe would be suitable for high-performance spectroscopy over a broad spectral range. Such features are never realized in one single facility, and concessions are always made. Most facilities use lasers for both the pump and probe sources. Many types of tunable pulsed lasers are available for the near-IR, visible, and near UV. Optical parametric oscillators (OPOs), which utilize laser mixing in nonlinear crystals, provide tunable, high-brightness, pulses spanning much of the infrared. However, they do not reach into the far-infrared region due to limitations imposed by absorption and phase matching in the non-linear mixing crystal. Their output can be parametrically amplified (OPA) to provide more power, but at much lower PRFs. Free-electron lasers (FELs) are moderately tunable and have a high power and brightness, but tend to have poor stability. Most have been built for the spectral range above 250 cm[-1] (wavelengths smaller than 40 μm). Femtosecond lasers can be used to produce coherent THz pulses through the rapid acceleration of photocarriers in a biased semiconductor. Such THz pulses are useful as the probe in a pump-probe experiment, and provide high spectroscopic signal-to-noise. However, because the rate of change in the generated photocurrent is limited to about 30 fs, the upper frequency limit for this method is about 100 cm[-1].[30] Recently, electro-optic rectification in thin chalcogenides crystals has been shown to increase the spectral range to above 1500 cm[-1], but the spectral coverage at lower frequencies in not complete.[31–33]

From the foregoing, we see that a common limitation of most pump-probe arrangements is a restricted spectral range. This restriction is particularly onerous if the system under study relaxes through a wide range of energies. Synchrotron radiation, a pulsed continuum source spanning the spectral regions from microwaves to x-rays, completely relieves the constraint on spectral range, although at the expense of temporal resolution. Whereas FELs, OPOs, and THz emitters can have sub-picosecond pulse widths, synchrotron pulse durations are typically from tens of picoseconds up to about 1 nanosecond, depending on the synchrotron facility and its mode of operation.

## II.3 Pump-probe and interferometry

Most infrared spectrometry is accomplished by interferometric techniques. In a Michelson interferometer, the incoming light is divided into two beams with a variable optical delay (or retardation) introduced into one of the beams by a movable mirror. The beams are recombined, producing interference, and detected. The detected signal as a function of (optical) path difference, known as an "interferogram," is the Fourier transform of the incident light power spectrum. Thus, an inverse Fourier transform of this signal yields the spectral content of the light. When used with a single element infrared detector, this Fourier-transform infrared (FTIR) spectrometer (or interferometer) has significant multiplex and throughput advantages over diffraction grating or prism spectrometers, allowing high signal to noise ratio (S/N) spectra to be acquired in short periods of time.

A considerable portion of the infrared flux reaching the detector is from the 300 K environment. Because this flux is approximately constant with time, FTIR spectrometers modulate the desired signal so that ac detection can be employed and the background signal rejected. There are two methods for achieving this modulation. One method, known as "rapid scan," moves the mirror at a constant velocity through the complete optical path difference range. A typical velocity has the optical path changing at about 1 cm/s. A broadband AC signal results, which for the mid-IR spectral range (600 cm[-1] to 4000 cm[-1]) spans 600 Hz to 4 kHz, compatible with typical infrared detector response times (bandwidth), low-noise amplifiers, and high-resolution digitizers.

The other method, referred to as step-scan, sequentially positions the movable mirror at discrete intervals, pausing at each step to acquire the signal (often integrated for a few seconds to improve S/N). A chopper modulates the beam signal, allowing lock-in detection methods. This technique is used in very far infrared spectrometers where it is difficult to move large mirrors at the velocities needed to bring modulation frequencies out from the detector's 1/f noise regime. Step-scan is also used in modulation-type experiments and a class of time-resolved measurements employing a fast (gated) detector.

The sample to be investigated is usually placed between the interferometer and the detector to minimize false signals from modulation of the sample's own thermal emission. This situation raises an interesting question for the case of a pulsed light source. The interferometer necessarily delays one of the beams in time with respect to the other, so that the light traveling toward the sample and detector consists of two pulses for every pulse incident on the beam splitter. If the delay is short the pulses overlap temporally, but if the delay is sufficiently long, the two pulses may not overlap. Although this situation might lead to concern about the interference process and the possible loss of time resolution, it should not. The interference process for a pulsed source is no different than for a continuous source. Consider a spectrometer system with a continuous, spectrally smooth source limited to the spectral range below a high frequency cutoff $\nu_0$. Assuming that the spectrometer optics do not impart any sharp spectral structure, the coherence length for this system is $\ell_0 = c/\nu_0$, which is the maximum optical path difference for which the interferometer will produce measurable interference. Placing an object with a narrow spectral feature into the optical system changes the coherence length. If the feature has a width $\delta\nu < \nu_0$, then the coherence length increases to $\ell = c/\delta\nu > \ell_0$ and observable interference occurs out to a greater optical path difference. In particular, $\ell$ is the interferometer path difference necessary to yield a spectral resolution of $\delta\nu$. Now consider a pulsed source with pulse length $L$. When $L > \ell$ (where $\ell = c/\delta\nu$ and $\delta\nu$ is the narrowest spectral feature to be detected) the situation is clearly the same as for the continuous source. The optical path difference needed to resolve the spectral feature is less than the pulse length, so the light from the two arms of the interferometer readily overlap in time, producing interference as required. But when $\ell$ exceeds $L$, it might seem impossible to achieve the interference necessary for resolving a narrow feature, since the two light pulses (produced by the interferometer), no longer overlap when brought back together. Fortunately this is not the case, since a proper treatment of the sample's response requires expressing the infrared pulse as the superposition of spectrally narrow Fourier components. These components normally cancel each other beyond the extent of the pulse, but when separated out by a sample having narrow ($\delta\nu$) spectral response, they persist over a length scale $\ell = c/\delta\nu$ that can be much greater than $L$. Therefore interference continues out to the optical path difference necessary for resolving a spectral feature even when it exceeds the pulse length. High-resolution spectroscopic measurements performed at the NSLS, at MAXLAB (Lund), and at LURE (Orsay) synchrotron facilities confirm the presence of interference under such circumstances.

The other concern is the potential loss in temporal resolution due to the time-separated pulses from the interferometer. Again, we consider a spectral feature having width $\delta\nu$. The transform-limited relaxation time for such a spectral feature is $\delta\tau = 1/\delta\nu$. This relaxation time sets the highest time resolution for which meaningful information can be obtained. To resolve such a feature, the interferogram must be sampled to an optical path difference of $\ell = c/\delta\nu$, imparting a time delay between the two probe pulses (split by the interferometer) of $\delta t = 1/\delta\nu$. Thus the uncertainty of the probe arrival time is the same as the transform-limited minimum relaxation time, and the loss of temporal resolution is insignificant.

Lastly, we note that the pump and probe repetition frequencies (PRFs) should be greater than the IR detector system bandwidth (and the FTIR spectrometer's modulation frequency). This restriction ensures that the arrival of a probe pulse, and not the instant of time when the detector is sampled, determines the delay time and is equivalent to having the detector sample (average) over many pump-probe cycles. Typical rapid-scan modulation frequencies are in the 100 kHz range, whereas PRF's for synchrotrons are tens of MHz, at least two orders of magnitude higher. Thus, in normal conditions, each position of the scanning mirror averages over a few hundred to a few thousands pump-probe events. Since most FTIR systems allow for a choice of modulation frequencies, it is not difficult to meet this requirement, even if a pump-probe experiment is performed at a very low repetition rate. The rather low modulation (chopping) frequency in a step-scan system avoids this problem in all practical situations.

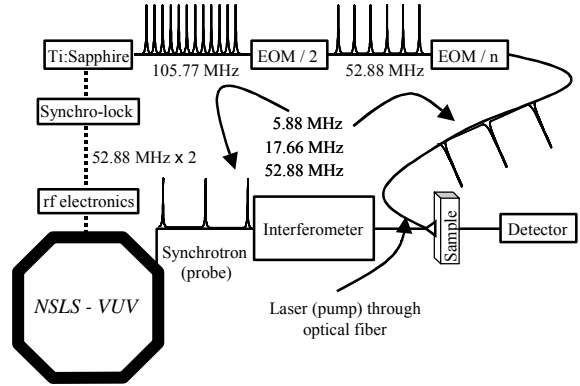

FIG. 2. Block diagram showing the pump-probe infrared facility at NSLS.

## III. INSTRUMENTATION

### III.1 Pump-probe pulses and synchronization

Fig. 2 shows schematically the pump-probe setup at NSLS. Our pump-probe apparatus utilizes a mode-locked Ti:sapphire laser as the excitation source and the synchrotron light as the probe. Here, we elucidate system details. The vacuum ultraviolet (VUV) ring at the NSLS operates with a 52.9 MHz RF accelerating system. Thus the minimum spacing between electron bunches is 18.9 ns. The revolution period (time for one full orbit) for a single bunch around the VUV ring's 51 m circumference is 170 ns, or 9 times the minimum bunch separation. Thus, the ring can simultaneously support a maximum of 9 equally spaced electron bunches, in which case synchrotron radiation is emitted at a 52.9 MHz PRF. Each position where a bunch of electrons can be maintained is referred to as an "RF bucket." The ring can be operated with specific RF buckets filled and others left empty. Symmetric bunch patterns, having filled buckets at equally spaced intervals, are convenient for synchronizing with a laser. The NSLS VUV ring has three such patterns; 1-bunch (a single bucket filled and the remaining 8 buckets empty), 3-bunch (every 3rd bucket filled), and 9 bunch, yielding PRFs of 5.88, 17.6 and 52.9 MHz respectively.[34] For normal operations, a second higher frequency RF system stretches the bunches (to increase the on-orbit lifetime of the electrons), resulting in a synchrotron emission pulse length in the nanosecond range. When this bunch stretching is disabled, the typical bunch length is between 500 ps and 1 ns. This second RF system can also be used in special operations to compress the electron bunches and achieve pulse durations of 400 ps or less. In general, the pulse length varies with electron beam current in the storage ring, but can be held reasonably constant over a modest range of currents by adjusting the degree of bunch compression from the second RF system. Additional reductions in bunch length can be attained through adjustments of the storage ring magnet system[35] or lowering the electron beam energy. The latter method reduces the high-energy cutoff for synchrotron radiation (in the soft x-ray region) and shifts the output wavelength for undulators, so is typically not employed. Presently, the shortest available synchrotron pulses, without seriously degrading overall storage ring operation, are near 300 ps. Efforts to reduce this value to 100 ps are in progress. Pulse widths for two operation modes of the VUV ring at NSLS are shown in Fig. 3. The pulses were detected with a Si photodiode and a 1 GHz digital oscilloscope (electronics raise/fall time of 400 ps) with the VUV ring in normal operations.

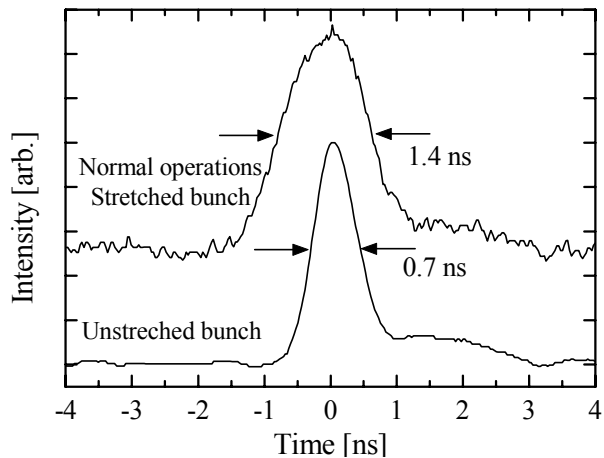

FIG. 3. Synchrotron pulse widths for normal operations with stretched bunches (top) and special rf cavity detuned operations that produce unstretched bunches (bottom). Obtained widths are an upper limit of their real value due to the 400 ps raise and fall time of the electronics used.

The pulsed pump source is a custom-engineered Ti:Sapphire mode-locked laser (Coherent Laser Group Mira 900P) producing 2 ps duration pulses at a repetition rate of 105.8 MHz — twice the VUV ring fundamental RF frequency (attempts to produce a longer cavity laser producing stable pulses at 52.9 MHz were unsuccessful.) The laser is tunable from 700 nm to 1000 nm and delivers an average power of nearly 1 W. A Coherent "Synchro-lock" system (also modified by the manufacturer for operation with a 52.9 MHz reference signal) is used to synchronize the laser's output pulse train to the VUV ring's pulses. For the latter, we derive a reference signal directly from the storage ring's RF cavity. This source of synchronization signal is preferable to the use of the RF system drive electronics because the phase relationship of the latter with the electron bunches may vary with the degree of beam loading. The synchro-lock system mixes the pulsed laser output signal with the reference signal (doubled to produce 105.8 MHz), producing an error signal that is used to correct the Mira cavity length. When the error signal is driven to zero, the laser is synchronized to the RF reference, in this case the VUV ring pulse train, but at a frequency of 105.8 MHz, twice the VUV ring PRF of 52.9 MHz. To reduce the laser repetition rate, a tuned electro-optic modulator (EOM, ConOptics model 360-40) rejects every other laser pulse. The laser power output is recovered by collecting the rejected pulse, delaying it by 9.45 ns (to fall in step with the next laser pulse in the stream), sending it through a ½-wave plate (to bring the polarization back to the original orientation) and inserting it back into the pulse stream. Measurements with a fast photodiode show that the pulses can be made coincident to within 10 ps.

A second "divide by n" EOM (ConOptics model 350-60) follows the first EOM to enable matching the laser PRF to the symmetric 3-bunch and 1-bunch patterns of the VUV ring. Thus the operating PRFs for our pump source are 52.9 MHz, 17.6 MHz and 5.88 MHz.

An adjustable delay between the pump and probe is necessary to carry out a time-resolved measurement. Since we desire delay times up to 170 ns, optical techniques are not practical. Instead, we employ an electronic delay on the laser, taking the VUV ring pulses as a fixed reference. In the present configuration, a voltage controlled phase shifter on the RF reference signal provides for a continuously adjustable delay up to 9.45 ns. Alternatively, a DC signal can be fed into the synchro-lock mixer circuit, providing a comparable range of pulse delay settings. Since the laser pulse period is 9.45 ns, larger delays can be obtained by switching the EOM to select a different pulse. Fig. 4(a) shows the laser pulses and their synchronization to the VUV ring light. Note that the normal multi-bunch operation of the VUV ring leaves two buckets empty.

This detail does not give rise to any detrimental effect in the pump-probe measurements, although two out of nine laser pulses are not useful for the measurement. Fig. 4(b) shows the pulse repetition rate from the laser (top curve) and in 9, 3, and 1 bunch operations modes when the laser beam passes through the EOM's.

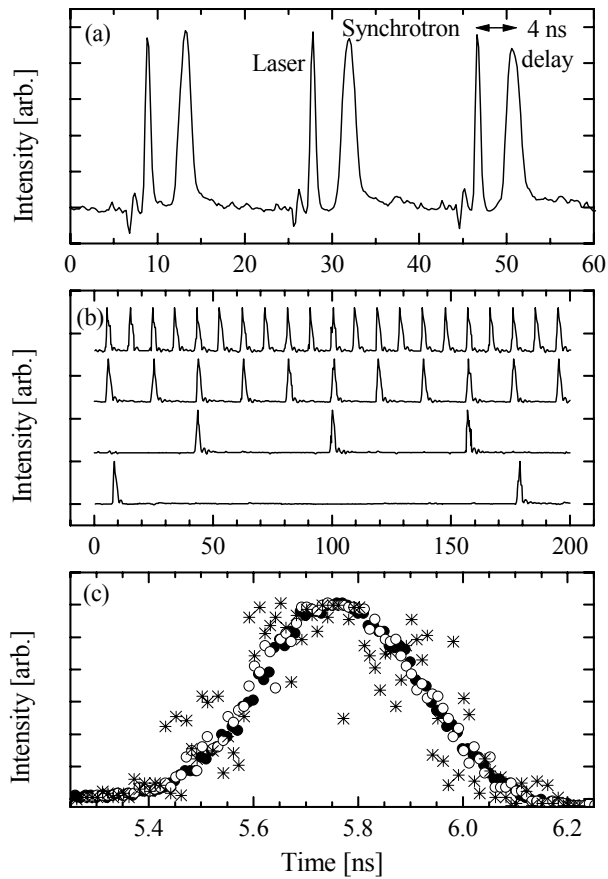

FIG. 4. (a) Synchronization between laser and synchrotron pulses for a 4 ns delay. As the synchrotron light also covers the near-infrared and visible part of the spectrum, the same detector can see both the laser and synchrotron pulses. (b) From top to bottom 105.9 MHz laser repetition rate; 52.88 MHz pulses obtained with divide-by-two EOM; and 17.63 and 5.88 MHz repetition rates obtained with simultaneous used of divide-by-two and divide-by-n EOM's. (c) The amount of scatter is a measure of the synchronization of laser pulses to the storage ring pulses. The stars correspond to using just the fundamental loop of the synchro-lock system. The open circles are for the harmonic loop, and the solid circles are for harmonic-loop plus high-bandwidth.

A final important issue to consider is the accuracy of the laser synchronization to the storage ring pulses. Fig. 4(c) shows the measured width and jitter of the laser pulses using the synchrotron RF cavity signal as a reference. The sampling oscilloscope requires many pulses to record data at each time point and produce the entire pulse shape. Timing jitter between the laser and the synchrotron reference can be detected as scatter along the horizontal (time) axis. The asterisks correspond to using just the fundamental loop of the synchro-lock system leading to a RMS jitter of at about 150 ps. The open circles are for the harmonic loop, and the solid circles are for harmonic-loop plus high-bandwidth where the jitter is well under 50 ps.

### III.2 Pulse delivery

Because the laser is physically separated from the infrared beamlines and their spectrometers, the laser light is transported to a given spectrometer using a 30 m length of multimode, graded-index optical fiber. An advantage of fiber delivery is the ease by which laser pulses can be fed to other beamlines at the NSLS. For 2 ps pulses at 800 nm (the fundamental of the Ti:Sapphire laser), the intrinsic dispersion due to the variation of the refractive index in glass, is 0.12 ps/m resulting in a pulse broadening of approximately 4 ps; this value does not require any dispersion compensation. Because our synchrotron pulses are not likely to be less than 100 ps long, this broadening can be neglected. The situation is different for a Ti:sapphire laser producing transform limited 100 fs pulses. In this case 30 m of single mode fiber broadens the pulse to about 30 ps. This broadening rises to about 80 ps if the pulses are frequency-doubled to produce light at 400 nm. In fact, as we use the laser with its picosecond optics, the temporal resolution of our complete system is determined entirely by the duration of the synchrotron pulse. Thus, we can take $I_{pump}(t'') = I_0 \, \delta(t'')$ and rewrite Eq. 1. as

$$S(\Delta t) = I_0 \int_{-\infty}^{+\infty} dt' I_{probe} (t' + \Delta t) \, G(t') \qquad (2)$$

showing that the measured response is a convolution of the synchrotron pulse shape with the sample's intrinsic response. The nature of damping in a storage ring leads to a Gaussian-shaped electron bunch and probe intensity profile.

Due to coupling losses, the output power from the optical fiber is reduced to approximately 70 % of the input value. The maximum average power that can be delivered to a specimen is about 500 mW, corresponding to 10 nJ/pulse, or about $4 \times 10^{10}$ photons/pulse. At the sample location, temporal coincidence between the pump and probe can be determined using a high-speed photodiode (150 ps rise time) and fast oscilloscope. Since the synchrotron output spans the near infrared and visible, a single detector can be used to monitor both pump and probe pulses. An example of this is shown in Fig. 4(a) for a pump-probe delay of 4 ns.

### III.3 Spectrometry: Infrared beamlines U12IR and U10A

Before describing the particular techniques used for collecting time-resolved spectra, we briefly describe two of the NSLS infrared beamlines (designed for solid state physics investigations) and their respective spectrometers. U10A extracts synchrotron radiation over the spectral range from about 50 cm⁻¹ up to 40000 cm⁻¹, and is instrumented with a Bruker IFS66v/S FTIR spectrometer that operates in either rapid scan or step scan modes. U12IR is capable of similar spectral coverage, but is optimized for long wavelengths. The originally installed spectrometers (a Bruker IFS113v and lamellar grating interferometer) have been replaced by a single step-scan instrument: a Sciencetech SPS-200 polarizing interferometer. The primary spectral range extends from about 500 cm$^{-1}$ down to about 2 cm$^{-1}$. Thus, between U12IR and U10A, time resolved spectroscopy can be performed for photon energies from 250 µeV up to several eV. In addition, both beamlines maintain the high-brightness of the synchrotron source, enabling small regions (limited only by diffraction) to be probed. Though most experiments are performed using pulse fluencies of less than 1 µJ/cm$^2$, an excitation fluence of 100 µJ/cm$^2$ per pulse can be achieved when the pump beam is focused to a 100 µm diameter spot, compatible with optimally focused synchrotron radiation out to wavelengths of 50 µm.

## IV. PHOTOINDUCED SPECTROSCOPIC METHODS

Pump-probe measurements can be performed by a variety of techniques. The basic approach is to set a particular pump-probe delay and then measure the specimen's spectral response (e.g., transmission or reflection) with the pump source "on" and "off"; the difference being the photoinduced signal, $\Delta T$ or $\Delta R$. Generally one plots $\Delta T / T$ or $\Delta R / R$ as a function of frequency. The transmission or reflection can of course be used to compute an intrinsic response function (such as the optical conductivity), which is now also a function of time (cf., $G(t)$ in Eq. 1). The measurement is repeated for other delay times until a complete picture of the decay has been built-up. As with conventional photoinduced spectroscopy, the challenge is to detect small spectroscopic changes, and high-sensitivity differential methods are employed whenever possible.

Because the particular changes of interest are those that occur as a function of time, we employ a differential technique that varies only the time difference between the pump and probe pulses. The pump-probe delay is modulated by a small amount $\delta t$ around a particular delay time $\Delta t$ using a sine or square wave signal. Thus $\delta S(\Delta t)$, the change in response $S(\Delta t)$ due to modulation $\delta t$, is given by the time derivative of Eq. 2.

$$\delta S(\Delta t) = I_0 \frac{\partial}{\partial t} \int_{-\infty}^{+\infty} dt' I_{probe} (t' + \Delta t) \, G(t') \delta t \qquad (3)$$

When the amplitude of the temporal modulation $\delta t$ is small, this measurement essentially gives the derivative of the photoinduced signal with respect to time.

In our setup, the synchro-lock electronics system provides a signal input for this purpose, but a voltage-controlled phase shifter on the RF reference signal works equally well. The modulation signal serves as a reference for lock-in detection, with the detected signal being the difference in the system's response between delay times $\Delta t_1$ and $\Delta t_2$. We typically use a temporal modulation of a few hundred picoseconds. In general, one may dither either the laser (pump) pulse or the infrared (probe) pulse, but varying the arrival times of electron bunches in the storage ring is not practical.

A key benefit of this differential technique is the rejection of false thermal signals. Though the properties of some mounted specimens can lead to thermal time constants of a few nanoseconds, longer thermal decay times (larger than 20 ns) are more common. In these situations, laser excitation using a 50 MHz PRF causes sample heating until a new equilibrium temperature is achieved. If the sample's optical properties are a function of temperature, then a measurement comparing the optical response between laser "off" and laser "on" will have a significant thermal contribution that may dominate over the photoexcitation signal of interest. In the differential technique, the laser is always illuminating the sample, maintaining a thermal steady state. The measurement is only sensitive to those optical properties changing on the ns time scale. This situation is in contrast to the case where the photoinduced signal is obtained by mechanically chopping the laser beam; the latter directly modulates the thermal signal as well.

Limitations in the response time of the laser system limit the dithering frequency to a few hundred Hz (depending on amplitude and choice of square wave or sinusoidal modulation). Such low frequencies are compatible with essentially all infrared detectors of interest (including high sensitivity bolometers for the far IR), but are not suitable for rapid-scan FTIR spectrometers. For this reason, step-scan interferometers (typical of most spectrometers for the very far IR[36]) are used in this differential technique. A different procedure is needed to perform a differential pump-probe measure with a rapid-scan interferometer. For this situation, a number of interferometer scans (one scan contains the full infrared spectral response) are collected and averaged for a particular pump-probe delay $\Delta t$. The time difference is set to $\Delta t + \delta t$ and another set of scans is taken. The delay is set back to $\Delta t$, and the entire procedure is repeated a number of times, resulting in an interleaved set of spectra for times $\Delta t$ and $\Delta t + \delta t$. The difference between the two sets of data is equivalent to the difference in the spectral response obtained by the lock-in differential measurement. The guiding principle is to prevent the various modulation frequencies from overlapping and mixing. The laser and synchrotron PRFs are greater than 5 MHz; well above the sampling frequencies for any interferometric technique. In the lock-in differential method, the dither modulation frequency is ~ 100Hz while the interferogram sampling frequency for a step-scan instrument is usually on the order of 1 Hz (or less). In the rapid scan method, the interferogram sampling frequency is between 10 Hz and

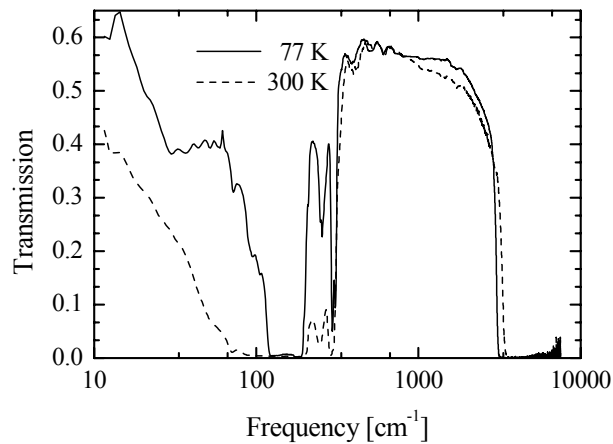

FIG. 5. Transmission of the MCT sample at 77 K (solid line) and 300 K (dashed line). The low frequency spectrum is dominated by the CdTe substrate phonons. At room temperature the band gap is located at about 3250 cm$^{-1}$ and it softens to 3000 cm$^{-1}$ at lower temperatures

100 kHz. By collecting interferogram scans over a minimum period of 5 seconds before switching to the other time delay, an effective dither modulation frequency below 0.1 Hz is achieved.

## V. ELECTRON-HOLE RECOMBINATION IN Hg$_{1-x}$Cd$_x$Te

As an example of the measurement capabilities of our system, we have investigated the dynamical response of Hg$_{1-x}$Cd$_x$Te (MCT). The infrared and electrical properties of this semiconductor, an important infrared detector material, have been widely investigated and are very well known.[37–39] In this experiment, we monitored the response of photogenerated free carriers in the MCT film. Pumping Hg$_{1-x}$Cd$_x$Te using light with photon energies greater than its bandgap creates mobile electrons and holes. Although photocarrier lifetimes for this material are nominally microseconds (for n-type) or hundreds of nanoseconds (for p-type), the lifetime can be much shorter in the presence of defects or a large surface recombination rate. Such photocarriers dynamics, in this and other semiconductors, is intrinsically related to the operation of fast infrared detectors.

### V.1. Experimental details

The sample consisted of an approximately 10 μm thick layer of MCT grown epitaxially on a Cd(Zn)Te substrate. The specimen studied had no special surface preparation (i.e., was unpassivated) and the surface had been left unprotected and exposed for some time; thus, a fast carrier relaxation could be expected due to a high surface recombination rate. At the laser wavelength (800 nm), MCT has a penetration depth of about 1 μm, much less than the MCT film thickness. Therefore, the laser creates no photo-carriers in the substrate. 300 mW of laser power was delivered to the sample. The far-infrared measurement (below 50 cm$^{-1}$) was performed in the lamellar grating step-scan spectrometer. The mid-infrared measurements (near the MCT band gap), employed the Bruker 113v fast-scan interferometer with Ge/KBr beamsplitter. The transmission of the sample was measured at several fixed temperatures between 4 K and 300 K in a flow cryostat. Polyethylene and KRS-5 windows were used for the far- and mid- infrared respectively. A liquid-He cooled silicon bolometer (response time around 0.5 ms) was used in the far infrared; a liquid nitrogen cooled InSb photodiode (response time of 1 μs including its amplifier) was used for the mid-infrared measurements. The measurements were taken during normal operations of the synchrotron ring.

The pump-probe measurement requires that the probe create a negligible excitation in the sample relative to the pump pulse. Since the synchrotron provides light over a very broad spectral range, the integrated intensity can be substantial (close to 50 mW across the entire infrared) and may create a significant population of excitations. Fortunately, the various optical elements (e.g. the interferometer's beamsplitter and vacuum windows) reduce the intensity and limit the spectral range sufficiently to alleviate this problem in most cases. For those where it does not, spectral bandpass filters can be used to constrain the spectral range to the region of interest and eliminate the problem.

To determine the overall time decay of the sample, we first do a spectrally integrated measurement. The interferometer modulator is left at a fixed position (preferably at the zero path difference — ZPD) and the detector signal is monitored for changes as the pump-to-probe delay time is varied around coincidence (zero relative delay time). The resulting signal represents the photoexcited response averaged across the entire spectral range of the experiment. From this we can determine the total time range and interval for collecting specific photo-induced spectra.

### V.2. Results

In Fig. 5 the sample transmission is shown at room temperature and at 77 K. Cooling the sample sharpens the phonons, reduces multi-phonon absorption processes and freezes out thermally generated carriers, all of which increase the low frequency transmission. The CdTe substrate phonons dominate the response between 100 and 400 cm$^{-1}$. The MCT band gap can be observed at room temperature at 3250 cm$^{-1}$, shifting to about 3000 cm$^{-1}$ when the sample is cooled.

Fig. 6 shows a typical measurement of the spectrally averaged response at a temperature of 100 K. The spectral range was limited to the far infrared (frequencies below 100 cm$^{-1}$) by the detector (a ℓHe cooled bolometer with integral filter) for which the absorption is linear in the free carrier density. Following Eq. 3, we integrated the differential signal to produce the response $S(\Delta t)$. This signal is a convolution of the intrinsic response $G(\Delta t)$ with the probe pulse shape. The latter can be described by a Gaussian and the former by a single exponential decay, i.e., $A \exp(-t/\tau_R)$. The solid line in Fig. 6 shows this convolution. The dashed line is the deconvolved decay. The full width of the Gaussian used is 960 ps, a consistent with the synchrotron bunch width when no bunch stretching is employed.

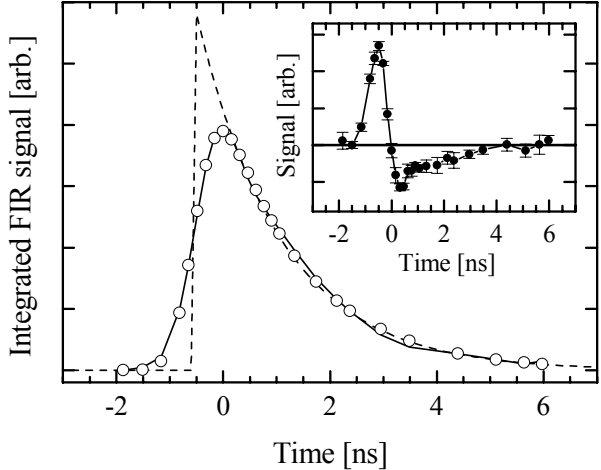

FIG. 6. Typical time resolved measurement. The inset shows the raw differential data with error bars. The main panel shows the signal integrated to produce the actual photo-induced transmission −ΔT/T (open circles) as a function of pump-probe delay time. The data was fitted (solid line) to an exponential decay, convolved with a 960 ps wide Gaussian that accounts for the probe pulse shape. The dashed line shows the exponential decay without the convolution.

In Fig. 7 we show the time dependence of the spectrally averaged signal for our sample at various temperatures as measured by the far-infrared bolometer detector. As noted above, this signal is proportional to the number of photo-carriers produced; obviously, it is the greatest when both pump and probe arrive at the sample simultaneously. A rapid decay in the nanosecond time range is observed. The points are the measured values and the solid lines are fits using the method described above. In the inset we show the values obtained for the recombination time $\tau_R$ as a function of temperature. This behavior is not unreasonable for n-type MCT with a substantial defect density.[10] The signal magnitude also varies with temperature due to substrate absorption.

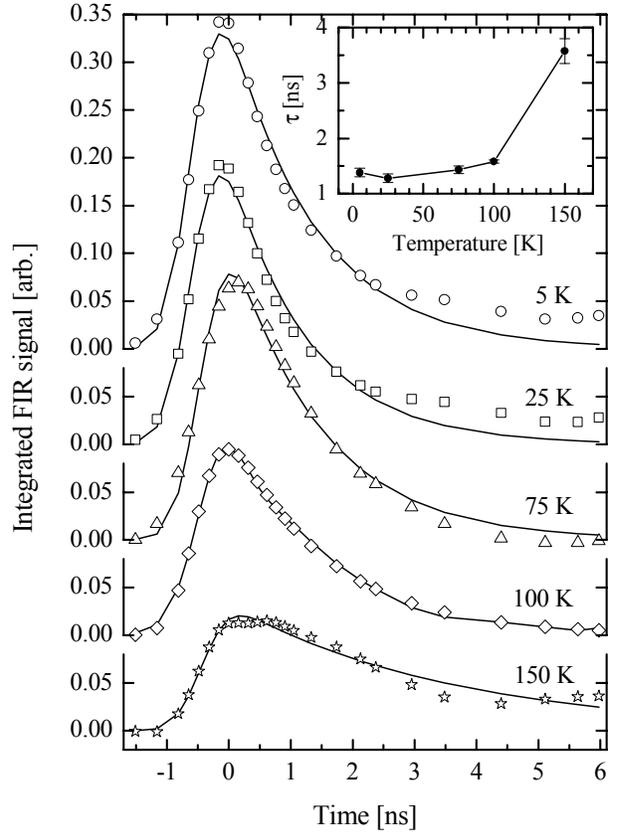

FIG. 7. Far IR absorption signal (spectrally integrated) for several temperatures. The solid lines are fits using a convolution of a Gaussian peak to a simple exponential decay. The inset shows the thermal evolution of the decay time. The full width at half maximum of the Gaussian in all fits is 0.96 ns.

Fig. 8 (a) and (b) show the photo-excited transmission change, $-\Delta T/T = -(T_{on} - T_{off})/T_{off}$ for various probe-to-pump delays in the far- and mid-infrared regions respectively. Data are shown in the frequency ranges where photo-induced effects were observed. Time dependent photo-induced signals are only observed when the laser is on and the pump-probe delay is less than 12 ns. No photo-induced signals are observed in the intermediate frequency range. Since the spectra were collected in a differential manner, the spectra shown in Fig. 8 were reconstructed by adding each differential spectrum to the previous one, beginning with the maximum pump-probe delay spectrum (for which the signal was zero). Fig. 8 (a) also shows a fit to the data using a Drude model described below.

### V.3. Discussion

The presence of the mobile electrons and holes can be detected by their absorption in the very far infrared. The electrons are more readily observed due to their smaller mass and longer mean free path (for phonon scattering). The absorption by free carriers manifests itself as an optical conductivity described by a Lorentzian function centered at zero frequency, the so-called Drude model

$$\sigma(\omega) = \frac{\sigma_0}{1 + (\omega\tau)^2} = \frac{\sigma_0}{1 + [\overline{\nu}/\Gamma]^2} \ , \qquad (4)$$

where $\sigma_0$ is the dc conductivity, $\omega$ is the angular frequency, and $\tau$ is the carrier scattering time. Eq. 4. also includes a more convenient expression in terms of wavenumber $\overline{\nu} = 1/\lambda$ and a Drude width $\Gamma = (2\pi c\tau)^{-1}$, both in units of cm$^{-1}$. The fits shown in Fig. 8 (a) all use the same $\Gamma = 16$ cm$^{-1}$ value for the scattering rate (a typical value for electrons in MCT). The only parameter that changes in these fits is $\sigma_0$, which incorporates the number of photo-excited pairs. That $\Gamma$ is quite small illustrates the need to work at very long wavelengths if one is to observe absorption by photo-excited carriers in this material.

optical response functions, known as the oscillator strength sum rule,[40] states that

$$\int_0^\infty \sigma_1(\omega)\,d\omega = \frac{\pi}{2}\frac{ne^2}{m} \ , \qquad (5)$$

where $\sigma_1(\omega)$ is the real part of the frequency dependent optical conductivity and $n$ is the number density of particles with charge $e$ and effective mass $m$. The optical conductivity is related to other response functions through

$$\sigma_1 = \varepsilon_v \varepsilon''\omega = 2\varepsilon_v n\kappa\omega \ , \qquad (6)$$

where $\varepsilon_v$ is the permittivity of free space, $\varepsilon''$ is the imaginary (absorptive) part of the material's dielectric response function, $\omega$ is the angular frequency of the light, and $n$ and $\kappa$ are respectively the real and imaginary parts of the material's complex refractive index.

Exposing a system to light cannot change the number

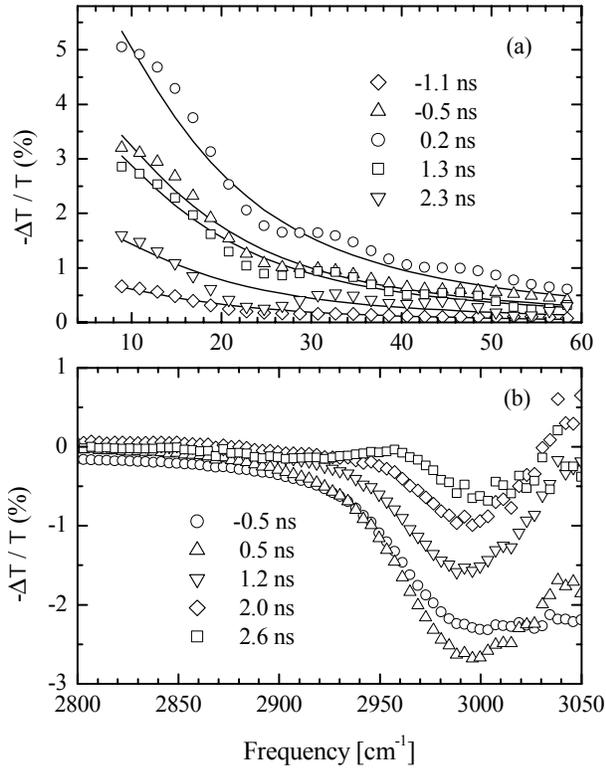

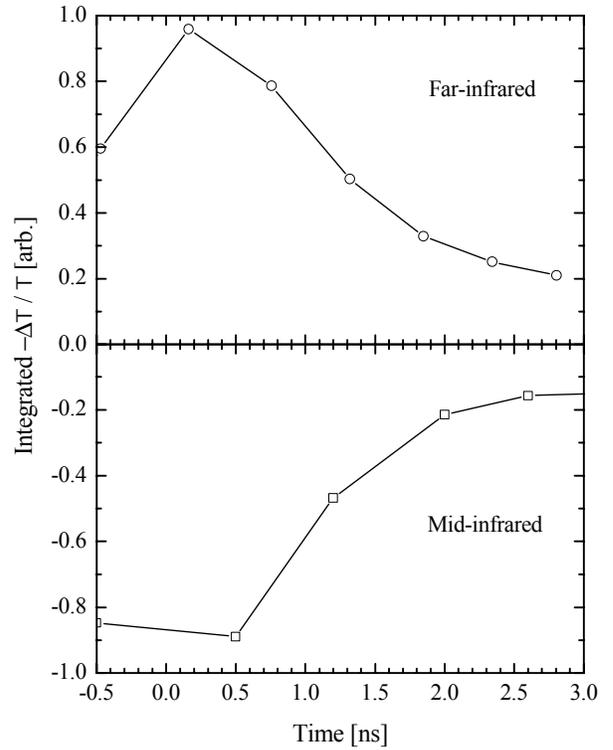

FIG. 8. Panel (a) shows the far-infrared time-resolved spectra of photo-excited excited MCT at 5K. The solid lines represent Drude curves with parameters (from top to bottom) $\Delta\sigma_0 = 0.07,\ 0.045,\ 0.04;\ 0.02;$ and $0.0085\ \Omega^{-1}\text{cm}^{-1}$. For all curves $\Gamma = 16$ cm$^{-1}$. (b) Time-resolved mid-infrared spectra near the MCT band gap at 5K. Here an absorption decrease (bleaching) is observed, in contrast to the absorption increase observed for the far-infrared. No response was observed for the spectral region in between.

FIG. 9. Integrated $-\Delta T/T$ in far- and mid-infrared. According to the sum rule the total integral under the $-\Delta T/T$ curve is a constant. The spectral weight created at low frequencies is subtracted from the band gap edge. The slight difference in the magnitude of two curves can be attributed to the different laser illuminated areas in the two measurements

The ability to cover a wide spectral range allows one to follow time-dependent processes that involve large shifts in optical absorption. An important sum rule that derives from the Kramers-Kronig relations for the of charged particles in it; thus a change in the absorption (conductivity) in one spectral region must be compensated by an opposite change elsewhere. In the limit of small transmission changes in a reasonably transparent material, the photo-induced signal (defined as $-\Delta T/T$) is proportional to the photo-induced change in the optical conductivity, $\Delta\sigma$. Thus the oscillator strength sum-rule implies that

$$\int_0^\infty \frac{-\Delta T(\omega)}{T(\omega)} d\omega = 0 \qquad (7)$$

This demonstrates that $-\Delta T/T$ must have an opposite behavior in another spectral range. In the case of a semiconductor, this opposite behavior occurs for frequencies near the interband absorption edge, which for this sample is near 3000 cm$^{-1}$.

This effect can be thought of as a time-dependent Burstein-Moss shift[41] (a shift of the interband absorption edge with carrier density). Finally, one can confirm that the oscillator strength sum rule holds as a function of time by integrating each $-\Delta T/T$ curve of Fig. 8. The results are shown in Fig. 9 for both the far and mid IR measurements. The two curves are indeed opposite in sign. We attribute the small difference (less than 10 %) in their magnitudes to unequal illumination intensity at the sample, which differed for the two spectrometers used in the measurements.

We have presented the characterization of a sub-nanosecond infrared time-resolved pump-probe facility using synchrotron light. Our facility can obtain time-resolved data as a function of temperature and frequency. A mode locked Ti:Sapphire tunable laser (700 – 1000 nm) is used as pump. The spectral range covered by the probe spectrometers spans 2 cm$^{-1}$ (0.25 meV) to 20 000 cm$^{-1}$ (2.5 eV). The capability to perform pump-probe measurements with either a step-scan or a fast-scan spectrometer has been demonstrated. A time resolution of 300 ps is achieved and a window for phenomena decaying as slowly as 170 ns is available at present. The time resolution in this technique is independent of the intrinsic detector response time. The facility is demonstrated in a measurement of photo-carrier relaxation dynamics for an MCT film, where decay times in the order of 1 ns were observed.

## ACKNOWLEDGEMENTS

This work was supported by DOE through contract DE-AC02-98CH10886 at the NSLS and DE-FG02-96ER45584 at the University of Florida.

[1] G. P. Williams, Nuc. Inst. and Meth. A **291**, 8 (1990).

[2] P. Roy, Y. L. Mathis, A. Gerschel, J. P. Marx, J. Michaut, B. Lagarde, and P. Calvani, Nucl. Instr. and Meth. Phys. Res. A **325**, 568 (1993).

[3] A. Marcelli, E. Burattini, C. Mencuccini, A. Nucara, P. Calvani, S. Lupi, and M. Sanchez del Rio, SPIE Accelerator-Based Infrared Sources and Applications Proc. **3153**, 21 (1997).

[4] R.P.S.M. Lobo, J.D. LaVeigne, D.H. Reitze, D.B. Tanner and G.L. Carr, Rev. Sci. Instrum. **70**, 2899 (1999).

[5] W. D. Duncan, and G. P. Williams, Appl. Opt. **22**, 2914 (1983).

[6] C. J. Hirschmugl, G. P. Williams, F. M. Hoffmann, and Y. J. Chabal, Phys. Rev. Lett. **65**, 480 (1990).

[7] G. L. Carr, J. A. Reffner, and G. P. Williams, Rev. Sci. Instrum. **66**, 1490 (1995).

[8] A. F. Goncharov, V. V. Struzhkin, M. S. Somayazulu, R. J. Hemley, and H. K. Mao, Science **273**, 218 (1996).

[9] J. Kircher, R. Henn, M. Cardona, P. L. Richards, and G. P. Williams, J. Opt. Soc. Am. B **14**, 705 (1997).

[10] G.L. Carr, J. Reichman, D. DiMarzio, M.B. Lee, D.L. Ederer, K.E. Miyano, D.R. Mueller, A. Vasilakis, and W.L. O'Brien, Semicond. Sci. Technol. **8**, 922 (1993).

[11] J.A. Mroczkowski, J.F. Shanley, M.B. Reine, P. LoVecchio, and D.L. Polla, Appl. Phys. Lett. **38**, 261 (1981).

[12] D. Y. Oberli, D. R. Wake, M. V. Klein, J. Klem, T. Henderson, and H. Morçok, Phys. Rev. Lett. **59**, 696 (1987).

[13] L. R. Testardi, Phys. Rev. B **4**, 2189 (1971)

[14] M. Johnson, Phys. Rev. Letters **67**, 374 (1991).

[15] C. J. Stevens, D. Smith, C. Chen, J. F. Ryan, B. Podobnik, D. Mihailovic, G. A. Wagner, and J. E. Evetts, Phys. Rev. Lett. **78**, 2212 (1997).

[16] B. J. Feenstra, J. Schützmann, D. van der Marel, R. Pérez Pinaya, and M. Decroux, Phys. Rev. Lett. **79**, 4890 (1997).

[17] D. Mihailovic, C. M. Foster, K. Voss, and A. J. Heeger, Phys. Rev. B **42**, 7989 (1990).

[18] J. Orenstein, and G. L. Baker, Phys. Rev. Lett. **49**, 1043 (1982).

[19] Z. Vardeny, J. Strait, D. Moses, T. C. Chung, and A. J. Heeger, Phys. Rev. Lett. **49**, 1657 (1982).

[20] G. P. Kelly, P. A. Leicester, F. Wilkinson, D. R. Worrall, L. F. Vieira-Ferreira, R. Chittock, W. Toner, Spectrochim. Acta A **46**, 975 (1990).

[21] J. H. Lim, O. V. Przhonska, S. Khodja, S. Yang, T. S. Ross, D. J. Hagan, E. W. Van Stryland, M. V. Bondar, U. Slominsky, Chem. Phys. **245**, 79 (1999).

[22] S. C. J. Meskers, P. A. van Hal, A. J. H. Spiering, J. C. Hummelen, A. F. G. van der Meer, R. A. J. Janssen, Phys. Rev. B **61**, 9917 (2000).

[23] G.L. Carr (*submitted*).

[24] G. L. Carr, J. Reichman, D. DiMarzio, M. B. Lee, D. L. Ederer, K. E. Miyano, D. R. Mueller, A. Vasilakis, and W. L. O'Brien, Semicon. Sci. Tech. **8**, 922 (1993).

[25] G. L. Carr, Vibrational Spectroscopy **19**, 53 (1999).

[26] G.L. Carr, R. P. S. M. Lobo, J. D. LaVeigne, D. H. Reitze and D. B. Tanner, Phys. Rev. Lett. **85**, 3001 (2000).

[27] G. L. Carr, M. Quijada, D. B. Tanner, C. J. Hirschmugl, G. P. Williams, S. Etemad, B. Dutta, F. DeRosa, A. Inam; T. Venkatesan, and X. Xi, Appl. Phys. Lett. **57**, 2725 (1990).

[28] W. Z. Shen, A. G. U. Perera, Infrared Phys. Tech. **39**, 329 (1998).

[29] See, for example "Ultrashort Laser Pulses: Generation and Applications", W. Kaiser, ed. (Springer-Verlag, Berlin, 1993).

[30] N. Katzenellenbogen, and D. Grischkowsky, Appl. Phys. Lett. **58**, 222 (1991)

[31] P.Y. Han, G. C. Cho, X. C. Zhang, J. Nonlinear Opt. Phys. Mat. **8**, 89 (1999).

[32] S. Kono, M. Tani, P. Gu, and K. Sakai, Appl. Phys. Lett. **77**, 4104 (2000).

[33] R. Huber, A. Brodschelm, F. Tauser, and A. Leitenstorfer, Appl. Phys. Lett. **76**, 3191 (2000).

[34] In fact the VUV ring normally operates with in a 7-bunch mode, giving the same time spacing as the 9-bunch mode (the pattern contains two empty bunches). This is not detrimental to pump-probe measurements, although two out of nine laser pulses are not useful.

[35] S.L. Kramer and J.B.Murphy, Proc. 1999 Particle Accelerator Conference **1**, 140 (1999).

[36] R. L. Henry, and D. B. Tanner, Infrared Phys. **19**, 163 (1979).

[37] M.W. Scott, J. Appl. Phys. **40**, 4077 (1969).

[38] E. Finkman, and Y. Nemirovsky, J. Appl. Phys. **50**, 4356 (1978).

[39] J.A. Mroczkowski, D.A. Nelson, R. Murosako, and P.H. Zimmerman, J. Vac. Sci. Technol. A **1**, 1756 (1983).

[40] See, *e.g.*, F. Wooten, Optical Properties of Solids – Academic Press, New York (1972).

[41] O. L. Doyle, J. A. Mroczkowski, J. A. Stanley, J. Vac. Sci. Technol. A **3**, 259 (1985).